\documentclass{article}

\begin{document}
\pagenumbering{arabic}

\title{Skyrmions and domain walls}
\author{B.M.A.G. Piette\thanks{e-mail address: B.M.A.G.Piette@durham.ac.uk},\\
W.J. Zakrzewski \thanks{e-mail address: W.J.Zakrzewski@durham.ac.uk} \\
Department of Mathematical Sciences\\University of Durham\\ Durham DH1 3LE,
UK}
\maketitle

\begin{abstract}
We study the $3+1$ dimensional Skyrme model with a mass term 
different from the usual one. We show that this new model possesses domain 
walls solutions. We describe how, in the equivalent 2+1 dimensional model, the 
Skyrmion is absorbed by the wall.
\end{abstract}

\def\p#1{\partial_#1}
\def\mod#1{ \vert #1 \vert }

\newcommand{\ee}{\end{equation}}
\newcommand{\be}{\begin{equation}}
\newcommand{\Ref}[1]{(\ref{#1})}

\section{Introduction.}
The Skyrme model has recently attracted a lot of attention as a reasonably 
good effective theory for describing nuclei \cite{Witten2}. It also plays
an important role in cosmology where the model describes cosmological 
structures\cite{ShellardVilenkin}. Moreover, on a purely mathematical ground, 
it is a good example of a model with stable topological solitons.

Introduced by Skyrme \cite{Skyrme}, the model was initially constructed 
as a model to describe the strong interactions. More recently, it
was shown to be a good effective model describing nuclei
in the limit of an infinite number of colours in QCD \cite{Witten}. 

The model is defined by the lagrangian
\be
{\cal L} 
= {f_{\pi}^2 \over 16} Tr\Bigl(\p{\mu}U \p{\mu}U^{\dag} \Bigr)
+ { 1 \over 32 e^2} 
  Tr \Bigl ( [(\p{\mu}U) U^{\dag}, (\p{\mu}U) U^{\dag}] \Bigr )^2
+ { f_\pi^2 m_{\pi}^2 \over 8} V(U),
\label{eLag}
\ee
where $U$ is an $SU(2)$ matrix, and where the potential is usually chosen as 
\be
V(U)= (Tr U - 2).
\label{eStdPot}
\ee

This potential term is responsible for making the pions 
of the model massive. The pion field $\bf\pi$ is related to the unitary field 
$U$ as follows: 
\be
U = exp \Bigl[ {2 i \over f_\pi} {\bf\pi} . {\bf \tau}\Bigr],
\label{eUpi}
\ee
and, when $U$ is close to the unit matrix, we have
\be
{f_{\pi}^2 m_{\pi}^2\over 8}(Tr(U)-2) \simeq 
          - {m_{\pi}^2 \over 2} \mod{\vec\pi}^2
\label{eSmallUPi}
\ee
showing that the potential \Ref{eStdPot} corresponds to a mass term for the 
pions. If instead of \Ref{eStdPot}, we choose
\be
V = {1\over 4} Tr(U)^2 - 1
\label{eNewPot}
\ee
for the potential, we have in the limit of small pion fields
\be
 {f_\pi^2 m_{\pi}^2 \over 8} ({1\over 4} Tr(U)^2 - 1) = 
   - {1\over 2} m_{\pi}^2\ \sin^2\mod{\vec\pi} \simeq  
   - {1\over2} m_{\pi}^2 \mod{\vec\pi}^2.
\label{eNewSmallPi}
\ee

We see thus that the two choices of potentials are identical in the limit
of small pion fields. On the other hand, the second choice has the property
of having 2 vacua: $U = \pm 1 $. 

\section{Domain wall solutions.}
It is convenient to describe the unitary field $U$ as an $S^3$ 
vector $\bf\phi$ defined by:
\be
U = \phi_0 1 + i \vec{\bf \tau}. \vec\phi
\label{eUPhi}
\ee
where $\phi_0 = \cos (2 \mod{\vec\pi} /f_{\pi})$ and 
$\vec\phi = \vec\pi/\mod{\vec\pi}\ \sin(2 \mod{\vec\pi}/f_\pi)$. 
Moreover, we have 
\be
V(U) = {1\over 4} Tr(U)^2 - 1 = (\phi_0^2 -1)
\label{eNewSmallPhi}
\ee
showing that the two vacua are given by $\phi_0 = \pm 1.$

The real vector $\phi = (\phi_0,\vec\phi)$ can be parametrised as follows
\be
\phi \, = \, \left( \matrix{ 
\cos f\ \cr 
\sin f\ \cos \Theta\ \cr 
\sin f\ \sin \Theta\ \cos \Omega\cr 
\sin f\ \sin \Theta\ \sin \Omega\cr} 
\right)
\label{ePhi}
\ee
where $f, \Theta$ and $\Omega$ are the usual polar angles on $S^3$. If we take 
for $\Theta$ and $\Omega$ any constant value, then $f$ is the only remaining 
field and the equation for the Skyrme model reduces to
\be
2\,f_{\mu\mu} + m_{\pi}^2 \sin (2 f ) = 0,
\label{eSinGEq}
\ee
which is the 3+1 dimensional sine-Gordon equation for the field
$2 f$ (one must remember that $f$ takes values in the interval $[0, \pi]$.)

If we restrict ourselves to solutions which depend only on one spatial 
variable, say $x$, then \Ref{eSinGEq} reduces to the usual 1+1 dimensional 
sine-Gordon equation and one of its solutions is the usual kink soliton:
\be
f = 2 \arctan[\exp(\pm m_{\pi} (x - x_0))],
\label{eKink}
\ee
where $x_0$ is a constant describing the position of the kink. 
Looking at \Ref{ePhi} we notice immediately that this solution  
extrapolates between the two vacua $\phi_0 = \pm 1$ of the model and
thus corresponds to a domain wall. 

Notice that the wall extends all the way to infinity and the boundary 
condition at infinity is different on each side of the wall. This means
that, strickly speaking, we cannot compactify the three dimensional Euclidean 
space into a sphere and so we loose the topology responsible for the
stability of the Skyrmion. One way to solve this problem is to close the
wall into a very large sphere. The curvature of the spherical wall would then
be so small that locally it would look like a flat planar wall.
Notice also that one can close the wall in various ways. The easiest
method consists in simply closing the wall into a sphere without changing
the fields. Another method would be to project the wall onto a sphere in such a 
way that the vector $\phi$ stays perpendicular to the wall. This is 
equivalent to identifying $\Theta$ and $\Omega$ with the polar angles on the 
sphere (instead of being constants); then the resulting wall carries one unit of
topological charge. This closed wall can also be thought off as a streched 
Skyrmion whose profile $f$ has been replaced by a kink-like function
centered at a very large distance from the origin and which corresponds to one 
of the two vacua inside the sphere. One would expect that such closed walls are
not stable but contract. In what follows we
will assume that the radius of the sphere is sufficiently large, that we can 
assume that the wall is flat and that is does not move.
  
When the wall is bent along the $x$ axis, the deformation propages as a 
wave along the wall.
To see this we seek a solution of the form $f = f_k(x,t) + d(x,y,z,t)$
where $f_k$ is the kink solution \Ref{eKink} and where $\mod{d} \ll 1$.
The linearised equation for $d$ is then given by
\be
d_{\mu\mu} + m_{\pi}^2 (1 - {2 \over \cosh^2(\mu x)})d = 0.
\label{eEqD}
\ee
It is a matter of straightforward algebra to show that 
\be
d = m_{\pi} {F(y,z,t) \over \cosh(m_{\pi} (x-x_0))} = 
m_{\pi} F(y,z,t) {\partial f_k \over \partial x} 
\label{eSolD}
\ee
is a solution of \Ref{eEqD} if $F$ satisfies the wave equation 
$F_{tt} - F_{yy} - F_{zz} = 0.$
The resulting solution corresponds to
a deformation of the wall along the $x$ axis which propagates according to 
the wave equation in the plane of the wall.

In \cite{TwoDPaper} a similar $2+1$ dimensional Skyrme model was studied.
The wall in this case reduces to a line and it was shown\cite{TwoDPaper} 
that the wall can carry deformation waves which propagate at the speed
of light. Moreover, the wall can also carry topologically charged
waves which also propagate at the speed of light. When a two-dimensional
Skyrmion is placed near the wall, the wall attracts the Skyrmion and
absorbs it. This produces 2 topological waves, each carrying half a unit of 
charge, which propagate in opposite directions.

It is also possible to reverse the process by colliding two topological waves 
combined with deformation waves. When they scatter, the two topological
waves produce a Skyrmion which then receives from the deformation waves 
enough energy to escape from the wall. This process is not very sensitive to
the initial conditions and thus seems to be relatively generic.

In the 3+1 dimensional case, we can expect a similar property to hold. 
This problem 
is still under investigation, but let us outline here a simple argument 
suggesting what is likely to happen. First of all, by looking at the field
configuration, it is easy to convince onself that, as in 2 dimensions, the 
wall can open up the Skyrmion and flatten it out on the wall. To describe this
it is convenient to use the cylindrical coordinates $r, \phi, z$. We choose 
$z$ as the axis perpendicular to the wall and passing through the center of 
the Skyrmion. $r$ and $\phi$ are the usual 2 dimensional polar coodinates on 
the wall and we fix the origin $r = z =0$ at the center of the wall. 
Because of the axial symmetry of the problem we see that, once the Skyrmion 
has flattened out on the wall, we have $\Omega = \phi$ and $f$ and $\Theta$ 
depend only on $z$ and $r$. We also see that $f$ will be given essentialy by 
the wall expression \Ref{eKink} plus some small corrections, while $\Theta$ 
will be $0$ at 
infinity and $\pi$ at the origin as it corresponds to a radial profile in
the plane of the wall. We can then compute the energy density of this 
configuration and perform a dilation of the radial variable $r$, leaving
$z$ unchanged. Doing this decreases the total energy of the configuration thus
suggesting that the Skyrmion will spread itself out on the wall. 

%\section{Conclusions}

\section{Acknowledgments}

We would like to thank A. Kudryavtsev for useful discussions.

%\refout 
 
\end{document}